\documentclass[12pt,a4paper]{article}

\title{}
\author{}
\date{}

\usepackage{amsmath}
\usepackage{amsfonts}
\usepackage{amssymb}
\usepackage{latexsym}

\makeatletter
\@addtoreset{equation}{section}
\makeatother

\newcommand{\be}{\begin{equation}}
\newcommand{\ee}{\end{equation}}
\newcommand{\ben}{\begin{equation}}
\newcommand{\een}{\end{equation}}
\newcommand{\bea}{\setlength\arraycolsep{2pt} \begin{eqnarray}}
\newcommand{\eea}{\end{eqnarray}}

\newcommand{\lbl}{\label}

\newcommand{\eq}[1]{(\ref{#1})}
\newcommand{\se}{\section}

\newcommand{\qd}{\quad}

\newcommand{\lt}{\left}
\newcommand{\rt}{\right}
\newcommand{\fr}{\frac}

\newcommand{\tf}{\tfrac}

\newcommand{\df}{\textrm{d}}

\newcommand{\pd}{\partial}

\newcommand{\sr}{\sqrt}

\newcommand{\ga}{\alpha}

\newcommand{\gc}{\gamma}
\newcommand{\gC}{\Gamma}

\newcommand{\gep}{\epsilon}

\newcommand{\gq}{\theta}

\newcommand{\gf}{\phi}

\newcommand{\gw}{\omega}

\newcommand{\bbR}{\mathbb{R}}

\newcommand{\ds}{\textrm{d} s^2}

\begin{document}


\thispagestyle{empty}

\begin{flushright}
MIFPA-11-10
\end{flushright}
\vspace*{100pt}
\begin{center}
{\bf \Large{A Dirichlet-type integral on spheres, applied to the fluid/gravity correspondence}}\\
\vspace{50pt}
\large{David D. K. Chow}
\end{center}

\begin{center}
\textit{George P. \& Cynthia W. Mitchell Institute for Fundamental Physics \& Astronomy,\\
Texas A\&M University, College Station, TX 77843-4242, USA}\\
{\tt chow@physics.tamu.edu}\\
\vspace{30pt}
{\bf Abstract\\}
\end{center}
We evaluate an analogue of an integral of Dirichlet over the sphere $S^D$, but with an integrand that is independent of $\lfloor (D + 1) / 2 \rfloor$ Killing coordinates.  As an application, we evaluate an integral that arises when comparing a conformal fluid on $S^D$ and black holes in $(D + 2)$-dimensional anti-de Sitter spacetime.

\newpage


\se{Introduction}


There is a class of functions that are particularly easy to integrate over the $n$-dimensional unit sphere $S^n$, namely monomials in the cartesian coordinates for $\bbR^{n + 1} \supset S^n$.  Let $x_i$, $i = 1 , \ldots , n + 1$ be such coordinates, so $S^n$ is the hypersurface $\sum_{i = 1}^{n + 1} x_i^2 = 1$.  A well-known result of Dirichlet \cite{dirichlet} is that, for non-negative integers $\ga_j$,
\ben
\int_{S^n} \! \prod_{j = 1}^{n + 1} x_j^{\ga_j} = \lt\{ 
\begin{array}{ll}
0 , & \textrm{some $\ga_j$ is odd} , \\ 
\dfrac{2 \prod_{j = 1}^{n + 1} \gC (\tf{1}{2} + \tf{1}{2} \ga_j)}{\gC (\tf{n + 1}{2} + \tf{1}{2} \sum_{i = 1}^{n + 1} \ga_i)} , & \textrm{all $\ga_j$ are even} .
\end{array}
 \rt.
\een
More generally, we have, for any real and non-negative $\ga_j$,
\ben
\int_{S^n} \! \prod_{j = 1}^{n + 1} | x_j |^{\ga_j} = \fr{2 \prod_{j = 1}^{n + 1} \gC (\tf{1}{2} + \tf{1}{2} \ga_j)}{\gC (\tf{n + 1}{2} + \tf{1}{2} \sum_{i = 1}^{n + 1} \ga_i)} .
\lbl{monomial}
\een
A simple direct proof is given in, for example, \cite{folland}.  For a historical review of a wider class of integrals, see \cite{gupric}.  Taking linear combinations of these results allows one to integrate polynomials and more general power series in $x_i$ over spheres.

In applications, it may be necessary to use some angular coordinates intrinsic to the sphere.  Consider the $D$-dimensional unit sphere $S^D$, and let $D = 2 n + \gep$, with $\gep = 0, 1$ according to whether $D$ is even or odd.  By introducing plane polar coordinates $(\mu_i , \gf_i)$ for orthogonal 2-planes in $\bbR^{D + 1}$, we have $\lfloor (D + 1) / 2 \rfloor$ angular coordinates $\gf_i$, $i = 1 , \ldots , n + \gep$, with independent periods $2 \pi$.  The flat metric on $\bbR^{D + 1}$ induces the round metric on $S^D$ given by
\ben
\ds_D = \sum_{i = 1}^{n + 1} \df \mu_i^2 + \sum_{i = 1}^{n + \gep} \mu_i^2 \, \df \gf_i^2 ,
\lbl{SDmetric}
\een
where $\mu_i$ satisfy the constraint
\ben
\sum_{i = 1}^{n + 1} \mu_i^2 = 1 .
\lbl{constraint}
\een
The metric coefficients are independent of $\gf_i$, i.e.~$\pd / \pd \gf_i$ are commuting Killing vectors; they represent rotational symmetries.  One can imagine situations in which one has to consider functions that are independent of $\gf_i$, and so are expressible in terms of $\mu_i$ only.  These are a generalization to higher dimensions of axisymmetric functions on $S^2$, which in 3-dimensional spherical polar coordinates depend on $\mu = \cos \gq$ but not the azimuthal coordinate $\gf$.  This motivates us to consider integrals that are analogous to \eq{monomial}, but over $S^D$ and involving powers of $\mu_i$.  The main result that we shall prove is that, for $\ga_j \geq - 1$,
\ben
\int_{S^D} \! \prod_{j = 1}^{n + \gep} \mu_j^{\ga_j} = \fr{2 \pi^{(D + 1)/2} \prod_{j = 1}^{n + \gep} \gC (1 + \tf{1}{2} \ga_j)}{\gC (\tf{D + 1}{2} + \tf{1}{2} \sum_{i = 1}^{n + \gep} \ga_i)} .
\lbl{main}
\een

As an application, we shall evaluate an integral arising in \cite{bhlalomi}, which concerns a correspondence between fluid mechanics on spheres and black holes in AdS (anti-de Sitter) spacetime.


\se{Proof of general result}


Let $X_I$, $I = 1 , \ldots , D + 1$ be cartesian coordinates for $\bbR^{D + 1}$.  We introduce sets of plane polar coordinates $(\mu_i , \gf_i)$ for the $(X_{2 i - 1}, X_{2 i})$-planes by
\ben
(X_{2 i - 1} , X_{2 i}) = (\mu_i \cos \gf_i , \mu_i \sin \gf_i) ,
\een
for $i = 1 , \ldots , n + \gep$.  If $D$ is even, then we instead define $\mu_{n + 1}$ by
\ben
X_{2 n + 1} = \mu_{n + 1} .
\een
The coordinates $(\mu_1 , \ldots , \mu_{n + 1} , \gf_1 , \ldots , \gf_{n + \gep})$ cover $\bbR^{D + 1}$, with ranges $\mu_i \geq 0$ for $i = 1 , \ldots , n + \gep$, $\mu_{n + 1}$ unrestricted if $D$ is even, and $0 \leq \gf_i < 2 \pi$ for all $i$.

The $D$-dimensional unit sphere $S^D \subset \bbR^{D + 1}$ is the hypersurface $\sum_{I = 1}^{D + 1} X_I^2 = 1$, on which the round metric is \eq{SDmetric}.  Bearing in mind the constraint \eq{constraint}, it can be expressed as
\ben
\ds_D = \ds_n + \sum_{i = 1}^{n + \gep} \mu_i^2 \, \df \gf_i^2 ,
\een
where
\ben
\ds_n = \sum_{i = 1}^{n + 1} \df \mu_i^2 .
\lbl{Snmetric}
\een

If we regard $\mu_i$ as cartesian coordinates for $\bbR^{n + 1}$, then \eq{Snmetric} can be interpreted as the round metric on $S^n \subset \bbR^{n + 1}$.  A difference is that there no constraints on the signs of $\mu_i$ as coordinates for $\bbR^{n + 1}$.  On the sphere $S^n$, we again have the constraint \eq{constraint}.

The interpretation of $\mu_i$ as either coordinates for $\bbR^{D + 1}$ or for $\bbR^{n + 1}$ enables us to reduce an integral over $S^D$ that is independent of the $\gf_i$ coordinates to an integral over $S^n$: we have a ``sphere within a sphere''.  Note that
\ben
\prod_{l = 0}^{n + \gep} \int_0^{2 \pi} \! \df \gf_l \, \int_{\sum_{i = 1}^{n + 1} \mu_i^2 = 1 , \, \mu_1 , \ldots , \mu_{n + \gep} \geq 0} \! \df^n \mu \, \prod_{j = 1}^{n + \gep} \mu_j^{\ga_j + 1} = \pi^{n + \gep} \int_{\sum_{i = 1}^{n + 1} \mu_i^2 = 1} \! \df^n \mu \, \prod_{j = 1}^{n + \gep} | \mu_j | ^{\ga_j + 1} ,
\een
because the $\gf_l$ integrals give a factor of $(2 \pi)^{n + \gep}$, and removing the sign constraints on $\mu_1 , \ldots , \mu_{n + \gep}$ gives a factor of $2^{- (n + \gep)}$.  The meaning of $\df^n \mu$ should be clear.  Explicitly, one can, for example, eliminate $\mu_{n + 1}$ from the integrand in favour of $\mu_1 , \ldots , \mu_n$ using the constraint \eq{constraint}.  Then $\df^n \mu$ means $\prod_{k = 1}^n \df \mu_k$, bearing in mind that for each choice of $(\mu_1 , \ldots , \mu_n)$ we must account for both signs of $\mu_{n + 1}$ on the right.  Expressing this in terms of integrals over $S^D$ and $S^n$, with respective metrics \eq{SDmetric} and \eq{Snmetric}, we have
\ben
\int_{S^D} \! \prod_{j = 1}^{n + \gep} \mu_j^{\ga_j} = \pi^{n + \gep} \int_{S^n} \! \prod_{j = 1}^{n + \gep} | \mu_j | ^{\ga_j + 1} .
\een
Using the Dirichlet integral \eq{monomial} for integration over $S^n$, remembering for even $D$ that it includes a factor of $\gC (\tf{1}{2}) = \sr{\pi}$, we hence obtain our main result \eq{main}.


\se{Application: fluid/gravity correspondence}


An explicit application of our main result is to a missing step in \cite{bhlalomi}, which studies the fluid/gravity correspondence.  It is argued that there is a duality between a conformal fluid on $S^D$ that solves the relativistic Navier--Stokes equations and a large black hole in AdS$_{D + 2}$ that solves the Einstein equations.  For one specific example, in arbitrary dimensions, the fluid is uncharged and rigidly rotating, and the black hole is the Kerr--AdS solution, with a horizon radius much larger than the AdS radius.  One can compare the thermodynamics of both sides of the correspondence.  From the correspondence for non-rotating solutions, one can make predictions for rotating solutions.

On the fluid side of the correspondence, one considers the spacetime
\ben
\ds = - \df t^2 + \ds_D ,
\een
where $\ds_D$ is the round metric on $S^D$ \eq{SDmetric}.  The spacetime is filled with a fluid with velocity
\ben
u^a \pd_a = \gc \bigg( \fr{\pd}{\pd t} + \sum_{i = 1}^{n + \gep} \gw_i \fr{\pd}{\pd \gf_i} \bigg) ,
\een
where
\ben
\gc = \fr{1}{\sr{1 - v^2}} , \qd v^2 = \sum_{i = 1}^{n + \gep} \mu_i^2 \gw_i^2 ,
\een
and $\gw_i^2 < 1$.  One computes the energy-momentum tensor and currents.  Integration gives conserved charges, which can be compared with the gravity side of the correspondence.  A missing step in \cite{bhlalomi} is a proof for all $D$ of a certain integral, namely
\ben
\int_{S^D} \! \bigg( 1 - \sum_{j = 1}^{n + \gep} \mu_j^2 \gw_j^2 \bigg) ^{- (D + 1) / 2} = \fr{2 \pi^{(D + 1)/2}}{\gC (\tf{D + 1}{2}) \prod_{j = 1}^{n + \gep} (1 - \gw_j^2)} .
\lbl{fluidintegral}
\een
Equivalently, we have
\ben
\int_{S^D} \! \gc^{D + 1} = \fr{V_D}{\prod_{j = 1}^{n + \gep} (1 - \gw_j^2)} , \qd V_D = \fr{2 \pi^{(D + 1)/2}}{\gC (\tf{D + 1}{2})} ,
\een
where $V_D$ is the volume of $S^D$.  Using this result, one then finds agreement between the two sides of the correspondence.

To prove the required integral, we first use binomial expansions to obtain
\ben
\bigg( 1 - \sum_{j = 1}^{n + \gep} \mu_j^2 \gw_j^2 \bigg) ^{- (D + 1) / 2} = \sum_{k_1 , \ldots , k_{n + \gep} \geq 0} (\tf{D + 1}{2}) (\tf{D + 1}{2} + 1) \ldots (\tf{D + 1}{2} + k - 1) \prod_{j = 1}^{n + \gep} \fr{(\mu_j \gw_j)^{2 k_j}}{(k_j)!} ,
\een
where
\ben
k = \sum_{j = 1}^{n + \gep} k_j .
\een
From the main result \eq{main}, if $k_j$ are non-negative integers, then
\ben
\int_{S^D} \prod_{j = 1}^{n + \gep} \mu_j^{2 k_j} = \fr{2 \pi^{(D + 1) /2} \prod_{j = 1}^{n + \gep} (k_j)!}{\gC (\tf{D + 1}{2} + k)} .
\een
Using this and the expansion of $\prod_{j = 1}^{n + \gep} (1 - \gw_j^2)^{-1}$, we hence obtain the integral \eq{fluidintegral}.





\begin{thebibliography}{99}

\bibitem{dirichlet}
Lejeune-Dirichlet, ``Sur une nouvelle m\'{e}thode pour la d\'{e}termination des int\'{e}grales multiples,'' \textit{J. Math. Pures Appl.} Ser. 1, {\bf 4}, 164 (1839).

\bibitem{folland}
G.B.~Folland, ``How to integrate a polynomial over a sphere,'' \textit{Amer. Math. Monthly} {\bf 108}, 446 (2001).

\bibitem{gupric}
R.D.~Gupta and D.St.P.~Richards, ``The history of the Dirichlet and Liouville distributions,'' \textit{Int. Stat. Rev.} {\bf 69}, 433 (2001).

\bibitem{bhlalomi}
  S.~Bhattacharyya, S.~Lahiri, R.~Loganayagam, S.~Minwalla,
  ``Large rotating AdS black holes from fluid mechanics,''
  JHEP {\bf 0809}, 054 (2008).
  [\texttt{arXiv:0708.1770}].

\end{thebibliography}
\end{document}